\documentclass{PoS}

\usepackage[utf8]{inputenc} 

\usepackage{amssymb}
\usepackage{fontenc}
\usepackage{times}
\usepackage{mathptmx}
\usepackage{graphicx}

\title{Spin period evolution of GX 1+4}

\ShortTitle{Spin period evolution of GX 1+4}

\author{\speaker{Ana González Galán}$^1$, Erik Kuulkers$^2$, Peter Kretschmar$^2$, Stefan Larsson$^3$, Konstantin Postnov$^4$, Al. Yu. Kochetkova$^4$, and Mark H. Finger$^{5,6}$\\
	\llap{$^1$} Department of Physics, Systems Engineering and Sign Theory, Alicante University\\
	P.O.~Box 99 E-03080 Alicante, Spain\\
	\llap{$^2$} European Space Agency, European Space Astronomy Centre\\
	 P.O.~Box 78, 28691, Villanueva de la Cañada, Madrid, Spain\\
	\llap{$^3$} Department of Astronomy, Stockholm University\\
	SE-106 91 Stockholm, Sweden\\
	\llap{$^4$} Sternberg Astronomical Institute\\
	119992, Moscow, Russia\\
	\llap{$^5$} National Space Science and Technology Center\\
	320 Sparkman Drive, Huntsville, AL 35805, USA\\
	\llap{$^6$} Universities Space Research Association\\
	6767 Old Madison Pike, Suite 450, Huntsville, AL 35806, USA\\
E-mail: \email{anagonzalez@ua.es}, \email{Erik.Kuulkers@sciops.esa.int}, \email{Peter.Kretschmar@esa.int}, \email{stefan@astro.su.se}, \email{kpostnov@gmail.com}, \email{sunny@sai.msu.ru}, \email{mark.h.finger@nasa.gov} }

\abstract{We report on the long-term evolution of the spin period of the symbiotic X-ray pulsar GX~1+4 and a possible interpretation within a model of quasi-spherical accretion. New period measurements from BeppoSAX/WFC, INTEGRAL/ISGRI and  Fermi/GBM observations have been combined with previously published data from four decades of observations. During the 1970's GX~1+4 was spinning  up with the fastest rate among the known X-ray pulsars at the time. In the mid 1980's it underwent a change during a period of low X-ray flux and started to spin down with a rate similar in magnitude to the previous spin up rate. The spin period has changed from $\sim$110~s to $\sim$160~s within the last three decades. Our results demonstrate that the overall spin down trend continues and is stronger than ever. We compare the observations with predictions from a model assuming quasi-spherical accretion from the slow wind of the M giant companion.\
          }

\FullConference{8th INTEGRAL Workshop “The Restless Gamma-ray Universe”\\
		September 27-30, 2010\\
		Dublin Castle, Dublin, Ireland}

\begin{document}

\section{Introduction}

Accreting X-ray pulsars are highly magnetized neutron stars in a binary system accreting matter from their companion star. The mass transfer can take place via Roche-Lobe overflow, strong stellar winds for giant stars or the Be emission mechanism. These accreting pulsars radiate predominantly in the X-ray band, and the radiation is modulated by the stellar rotation of the pulsar. For a review on this subject, see e.g.,~\cite{nagase89}.

GX~1+4 is an accreting X-ray pulsar discovered in 1970 by a balloon X-ray observation at energies above 15 keV showing pulsations with a period of about two minutes (\cite{lewin71}). It was one of the brightest X-ray sources in the Galactic center region. The composite emission spectrum of GX~1+4 indicated that the object was almost certainly a binary system, consisting of a symbiotic red giant and a much hotter source (\cite{davidsen77}). \cite{roche97} confirmed the optical companion to be the M giant V2116 Oph.

More recently, observations have shown a well-determined, 1161 day period, single-line spectroscopic binary orbit (\cite{hinkle2006}). The inclination of the orbit is, however, unknown. Therefore, on the basis of similar characteristics seen in other X-ray binary systems the mass of the neutron star is inferred to be $\sim1.35$ $M_\odot$. The mass function of the system, combined with the abundances of the M giant star and the assumed mass for the neutron star, indicate a mass of $\sim1.2$ $M_\odot$ for the M giant star. This implies that the M giant star is a first ascent giant which does not fill its Roche Lobe. In consequence, the neutron star is capturing the slow stellar wind of its companion, which makes V2116 Oph quite different from other low-mass X-ray binaries  (\cite{hinkle2006}).

Currently there is no accurate estimation of the magnetic field of GX~1+4. Assuming the standard accretion disk theory, the magnetic field has been estimated to be $B>3.7\times10^{13}$G (\cite{cui2004}), which is among the largest measured for accreting X-ray pulsars. However, there have been detected some signatures of the existence of CRSFs in the X-ray spectra of the neutron star which points to a value of $B\sim10^{12}$G for the magnetic field (\cite{ferrigno2007}).

Therein, GX~1+4, is a unique pulsating X-ray binary in many aspects. Among its singularities, it is the first low mass X-ray binary with an M giant optical companion, the first symbiotic X-ray pulsar ever observed, could have the highest magnetic field of any known accreting X-ray pulsar and has a peculiar spin behavior which explanation is the aim of this work.

\section{Results}

\begin{figure}
\center
\includegraphics[width=0.8\textwidth]{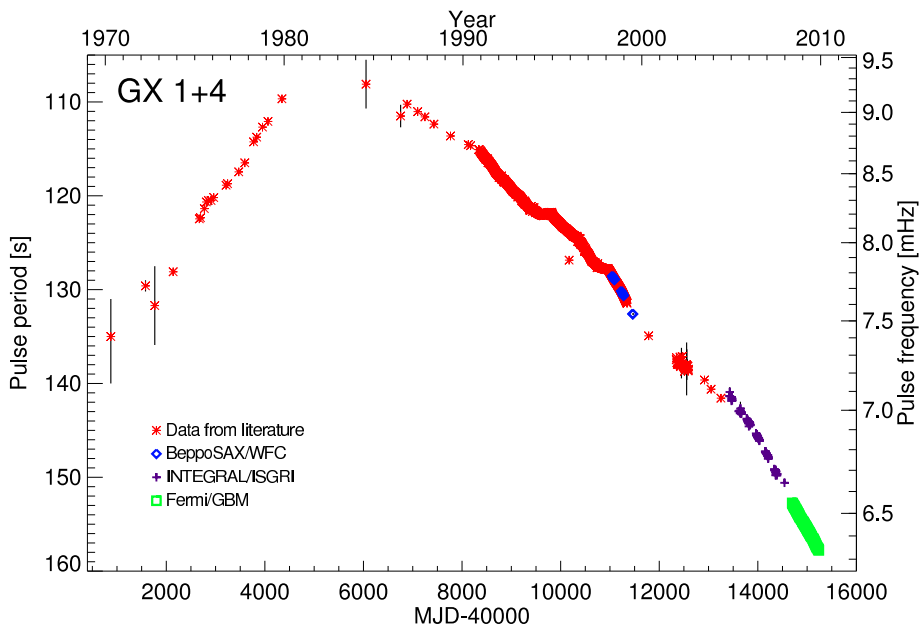} 
\caption{\label{GXhistdata}Long-term pulse period of GX~1+4 (\cite{gonzalez2010} and references therein) spanning a time of about 40 years. Note that the period increases from top to bottom.}
\end{figure}

\begin{figure}
\center
\includegraphics[width=0.45\textwidth,angle=180]{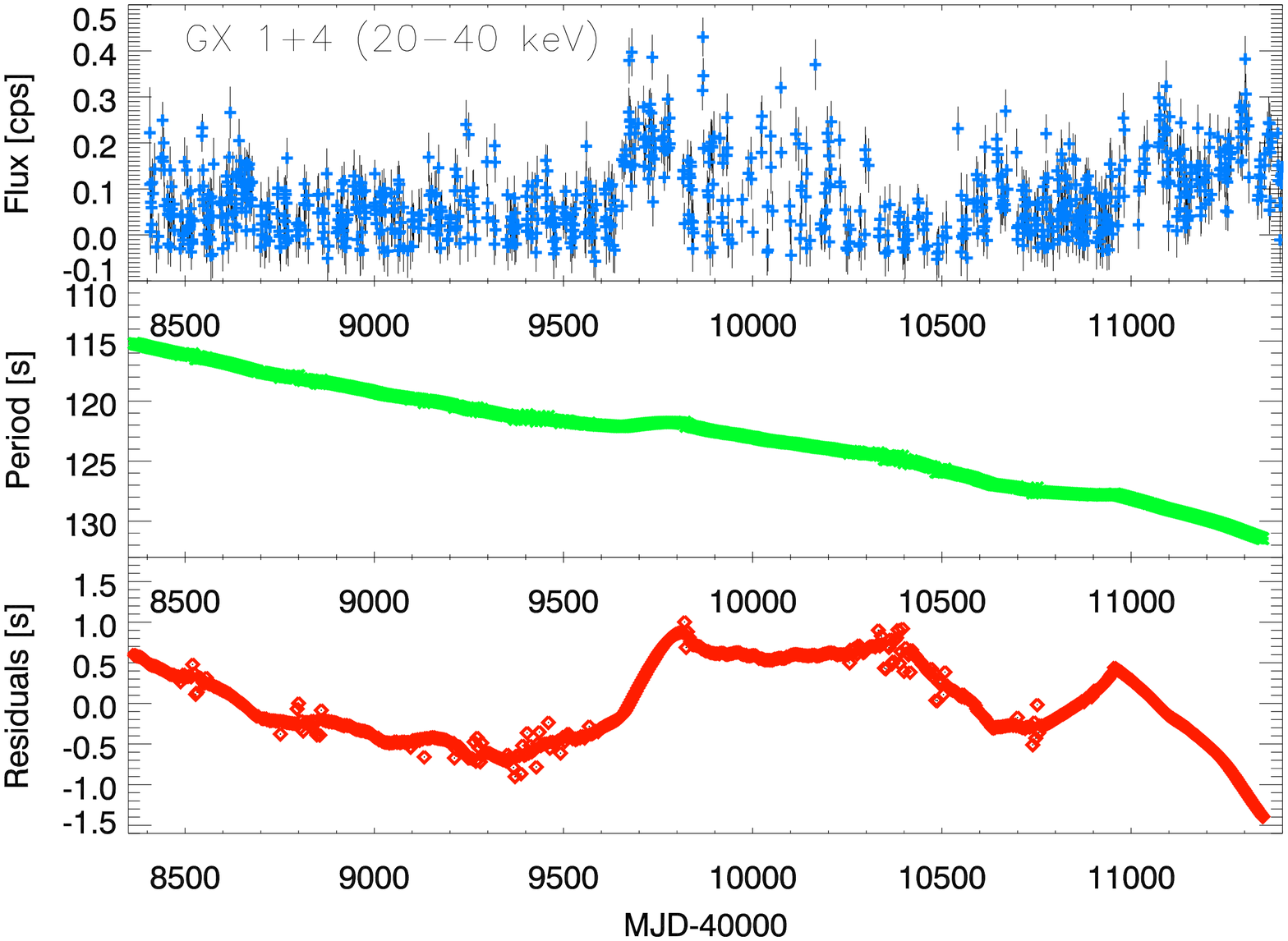}~
\includegraphics[width=0.45\textwidth,angle=180]{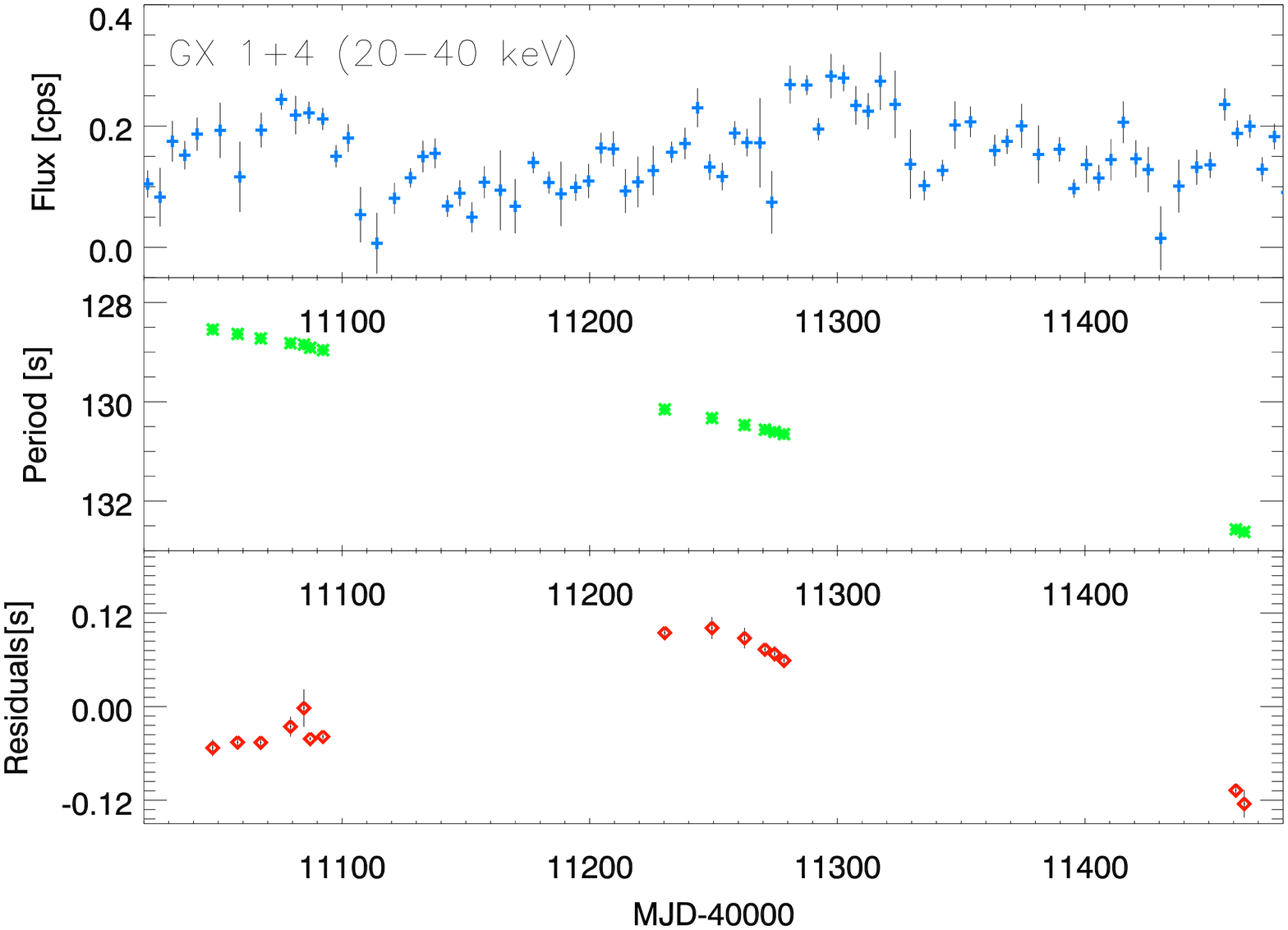} \\
\includegraphics[width=0.45\textwidth,angle=180]{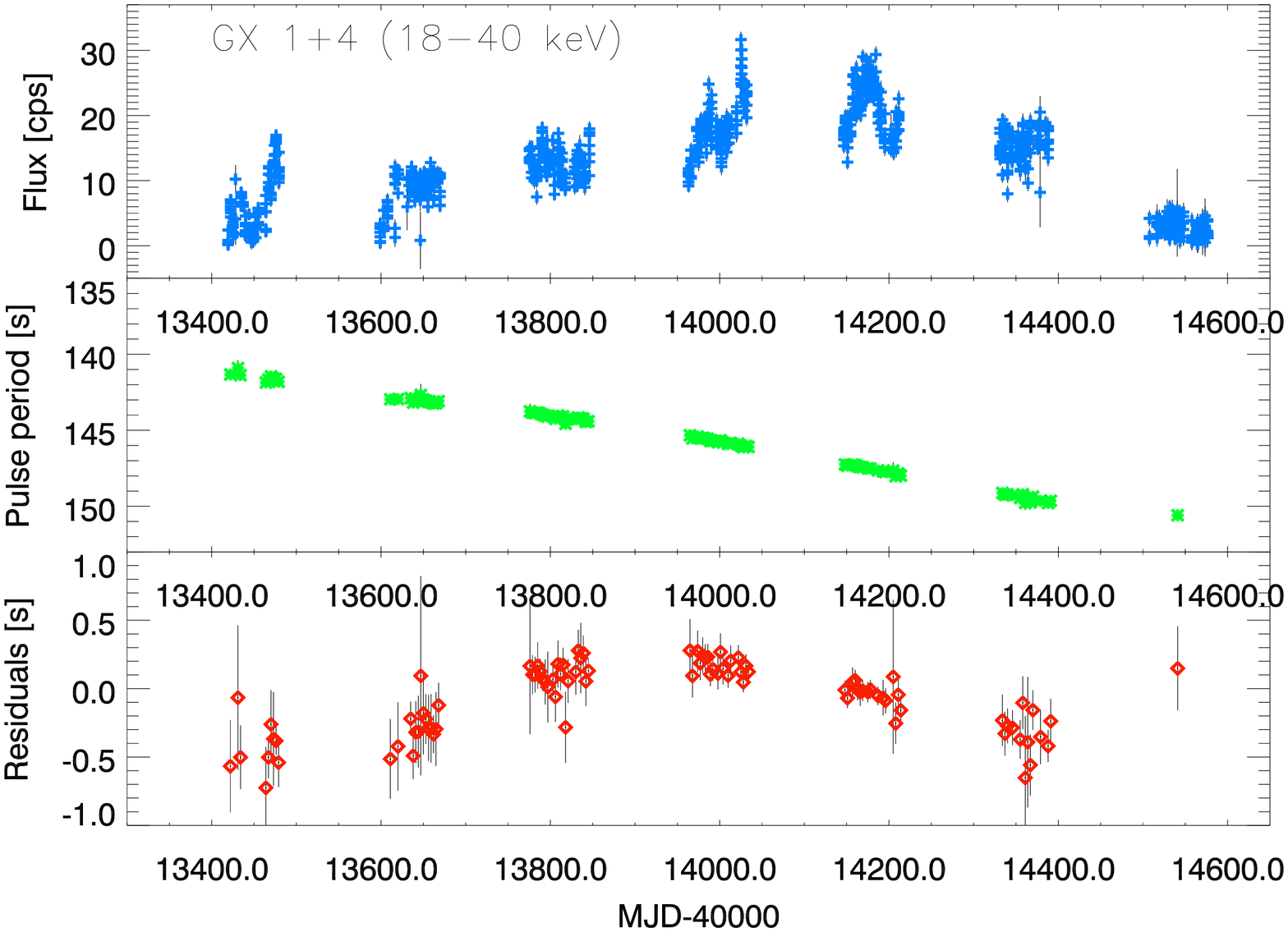} ~
\includegraphics[width=0.45\textwidth,angle=180]{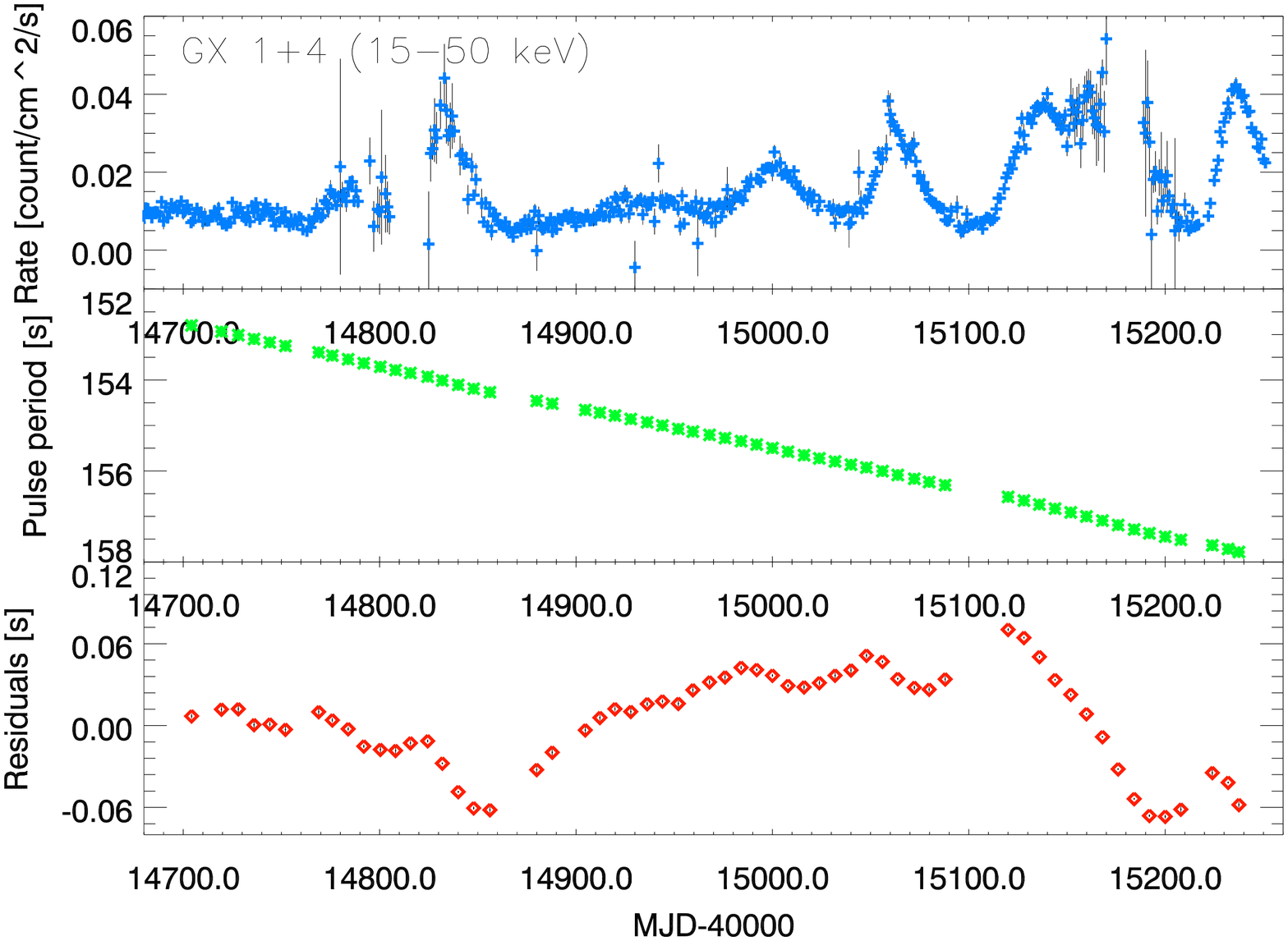} 
\caption{\label{fluxesperiods} GX~1+4 data. \scriptsize{{\bf Top left plot:} {\it Top:} CGRO/BATSE daily lightcurve in the energy range 20-40 keV. {\it Middle:} Pulse periods derived from CGRO/BATSE data. {\it Bottom:} Residuals of the periods from a linear fit. {\bf Top right plot:} {\it Top:} CGRO/BATSE daily lightcurve in the energy range 20-40 keV. {\it Middle:} pulse periods derived from BeppoSAX/WFC data. {\it Bottom:} Residuals of the periods from a linear fit. {\bf Bottom left plot:} INTEGRAL/ISGRI data. {\it Top:} Average flux per pointing in the energy range 18 to 40 keV. {\it Middle:} 20-40 keV pulse period. {\it Bottom:} Residuals of the periods from a linear fit. {\bf Bottom right plot:} {\it Top:} Swift/BAT daily averaged light curve in the energy range 15--50 keV. {\it Middle:} Pulse periods derived from Fermi/GBM data. {\it Bottom:} Residuals of the periods from a linear fit. Note that in all these plots the period increases from top to bottom.}
}
\end{figure}

The main result obtained during this work is the plot shown in Fig.~\ref{GXhistdata} which currently is the most updated and complete series of pulse periods of GX~1+4. The available measurements of the pulse period of the neutron star of GX~1+4 span a time period of about 40 years giving us a unique insight in the pulse period evolution of this symbiotic X-ray pulsar. To obtain this result we have combined all measurements that we are aware of from the literature with the pulse periods determined from BeppoSAX/WFC, INTEGRAL/ISGRI and Fermi/GBM data.

The system has a peculiar spin history. During the 1970's GX~1+4 was spinning up with the fastest rate ($\dot{P}\sim7.23 \times 10^{-8}$~s/s) among the known X-ray pulsars at the time (see e.g., \cite{nagase89}). During that period the source had an X-ray luminosity between $9 \times 10^{38}$ and $2.25 \times 10^{38}$~erg~s$^{-1}$ (\cite{white83}) (the luminosity value has been corrected to the distance range (3-6~kpc) (\cite{roche97}) ). However, EXOSAT observations in 1983 revealed an extended low state with an X-ray luminosity of $<$~$4 \times 10^{36}$~erg~s$^{-1}$ (\cite{hall83}). Around the same time GX~1+4 started to spin down with a rate similar in magnitude to the previous spin-up rate (see Fig.~\ref{GXhistdata}) and continues spinning down up to date. Indeed, the spin-down rate measured by Fermi/GBM is the strongest spin-down rate observed to date ($\dot{P}\sim1.08 \times 10^{-7}$~s/s), and the pulse period of the pulsar has increased by about 50\% during the last $\sim30$ years to the largest value ever known for this source ($\sim160$~s) (\cite{gonzalez2010}). 

Apart from the long-term plot, looking at the behavior along shorter time scales (Fig.~\ref{fluxesperiods}), it is possible to observe that even when the X-ray flux has variations, pulse period evolution seems to be almost linear.

The unusual long-term spin behavior of GX~1+4 has attracted considerable interest for many years. Studying the pulse period evolution in an accreting X-ray pulsar and relating it to, e.g., the luminosity changes, allows to test models of accretion and to gain insights on the physical processes taking place in this kind of systems. In \cite{gonzalez2010} we extend the investigation of the spin period history of GX~1+4 with new observations obtained by BeppoSAX/WFC, INTEGRAL/ISGRI and Fermi/GBM, we correlate X-ray flux variations with pulse period variations of the system using the same observations as well as those obtained with Swift/BAT and CGRO/BATSE, and provide a likely explanation for the spin history seen.

\section{Accretion models}
\label{theoretical}

Standard accretion theory by \cite{ghoshlamb79} assumes the formation of a prograde disk around the neutron star. Therefore, spin down of the pulsar is only possible in the quasi-equilibrium state, which implies the unusual magnetic field of $B\sim10^{13}$G. This prograde disk accelerates the neutron star through the angular momentum of the accreted matter. Thus, there is a positive correlation between accretion rate ($\dot{M}$) and the acceleration of the neutron star. Assuming $\dot{M} \propto F_{X}$, the positive correlation $\dot{\nu} \propto F^{6/7}_{X} $ (\cite{ghoshlamb79}) should be observed in the data even during spin down.

Transient disks with an alternating sense of rotation are known to form in numerical simulations in binary systems fed from stellar wind (\cite{frixell88}), hence, it is also possible the formation of a transient retrograde disk in GX~1+4 as it is a wind-fed source. Indeed, a retrograde accretion disk was proposed for the first time by \cite{makishima88} to explain the spin down observed in the neutron star of GX~1+4. This model assumes a retrograde disk around the neutron star decelerating the pulsar through the angular momentum of accreted matter, therein, the negative correlation $-\dot{\nu} \propto F^{6/7}_{X}$ should be observed in the data. Within this scenario the pulsar would necessarily be far from its equilibrium period. Therefore, a magnetic field of $B\sim10^{12}$G, instead of the high magnetic field predicted by standard accretion disk ($B\sim10^{13}$G), is required.

In \cite{gonzalez2010} we propose, for the first time, a third model to explain the behavior of GX~1+4, the quasi-spherical accretion model. Instead of a disk, this model predicts a quasi-static atmosphere around the neutron star which carries away angular momentum from the neutron star magnetosphere but allows accretion (in contrast to the so-called subsonic propeller regime (\cite{davies80})). As in the retrograde disk accretion model, this quasi-spherical scenario also predicts a  negative correlation between accretion rate and the acceleration of the neutron star during spin down, but within this scenario this correlation is weaker: $-\dot{\nu} \propto F^{3/7}_{X}$ (\cite{shakura2010}; \cite{postnov2010}). The formation of this quasi-static atmosphere is only possible for wind-fed sources like GX~1+4. This model does not require a magnetic field of $B\sim10^{13}$G as the pulsar does not need to be near its equilibrium period, but does not discard this unusual high magnetic field.

\section{Discussion}

\begin{figure}
\center
\includegraphics[width=0.7\textwidth]{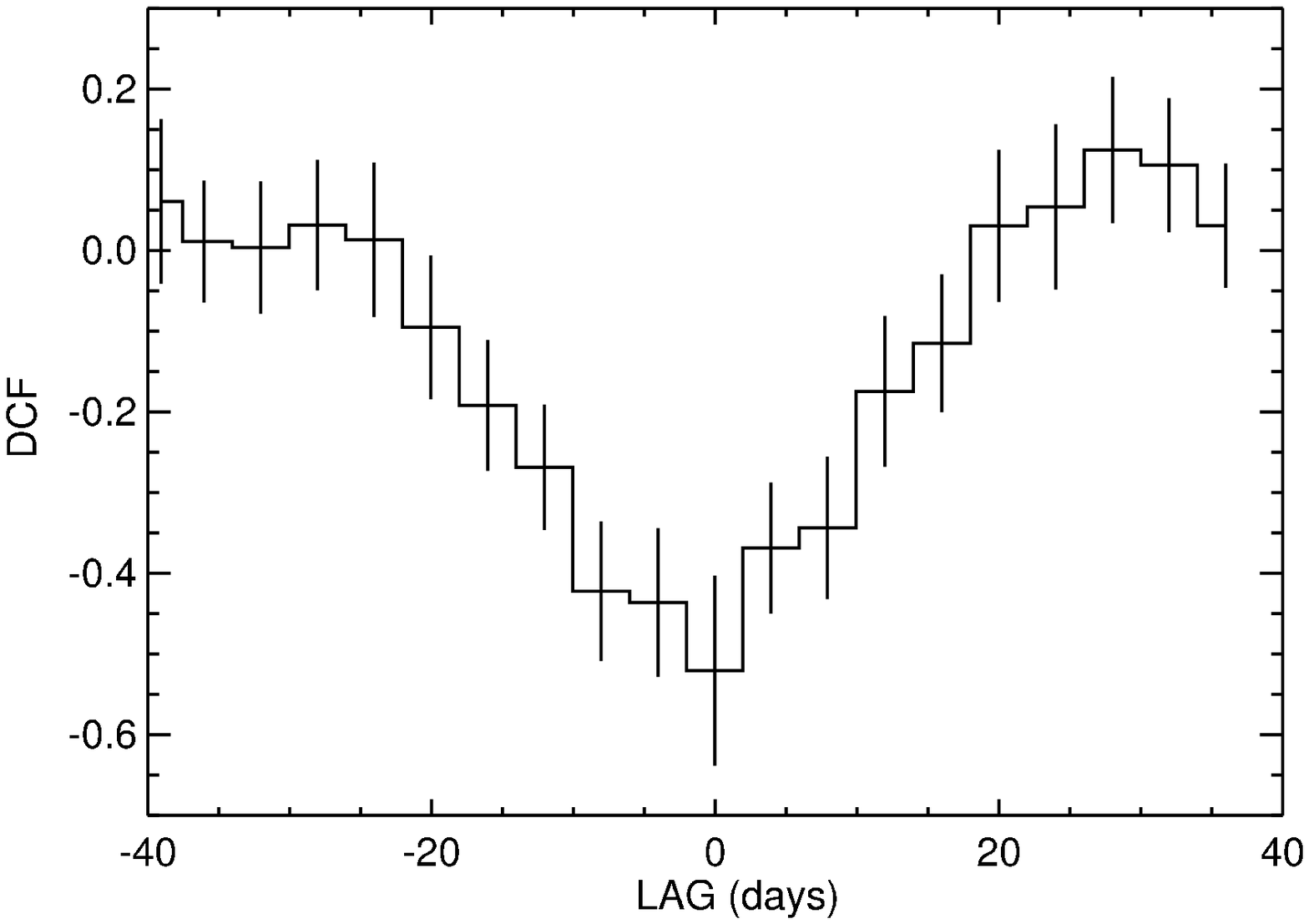}
\caption{\label{crosscorr}The discrete correlation function between $\dot \nu$ obtained from Fermi/GBM GX~1+4 data and the Swift/BAT 15 - 50 keV X-ray flux shown in bottom right panel of Fig.~2. The minimum near zero lag implies a negative correlation between the pulse frequency derivative and the X-ray flux (\cite{gonzalez2010}).}
\end{figure}

The correlations between X-ray flux and pulse period evolution predicted by the models proposed in Section~\ref{theoretical} can be tested by comparison with a regular sequence of X-ray fluxes and pulse periods which can be found in CGRO/BATSE and Fermi/GBM data (see Fig.~\ref{fluxesperiods}) for GX~1+4.

A cross-correlation of $\dot\nu$ obtained from Fermi/GBM data and the contemporaneous X-ray flux measured by Swift/BAT finds a minimum near zero lag (see Fig.~\ref{crosscorr}) implying a negative correlation between this two parameters for GX~1+4. Furthermore, this negative correlation has already been found previously in CGRO/BATSE data (i.e. \cite{chakrabarty97}). As the standard accretion disk model predicts a positive correlation in every case, this negative correlation discards the standard accretion disk model as a possible interpretation of GX~1+4 behavior.

Due to several reasons, including this negative correlation, different authors have tried to explain GX~1+4 pulse period behavior assuming the formation of a retrograde disk around the neutron star by matter captured from stellar wind of the secondary M-giant star in GX~1+4 (e.g. \cite{makishima88}, \cite{dotani89}, \cite{chakrabarty97}, \cite{nelson97}, \cite{ferrigno2007}). This model predicts a strong negative correlation between mass accretion rate and the acceleration of the pulsar ($-\dot{\nu} \propto F^{6/7}_{X}$). The correlation found between Fermi/GBM pulse periods and Swift/BAT X-ray flux is $-\dot{\nu} \propto F^{\sim0.30}_{X}$ (\cite{gonzalez2010}). Therein, the correlation found has the same sign as predicted by retrograde disk accretion model, but is about 3 times weaker than predicted. Indeed, the negative correlation found in CGRO/BATSE data ($-\dot{\nu} \propto F^{\sim0.48}_{X}$ (\cite{chakrabarty97})) is also weaker than predicted by this model. Apart from this, the weak points of the retrograde disk interpretation, include the puzzling long-term ($\sim30$ years) stability of such a retrograde disk.

This discrepancies between predicted and observed correlations and the puzzling long-term stability of the retrograde disk interpretation has led us to a new model, the quasi-spherical accretion model. Whithin this model, there is no such a difference between predicted ($-\dot{\nu} \propto F^{3/7}_{X}$) and observed ($-\dot{\nu} \propto F^{\sim0.30}_{X}$(\cite{gonzalez2010}), $-\dot{\nu} \propto F^{\sim0.48}_{X}$ (\cite{chakrabarty97})) negative correlation, and it is possible to explain the observed correlation without the formation and maintenance, during such a long time period ($\sim 30$ years), of a retrograde accretion disk (\cite{gonzalez2010}).

\section{Conclusions}

New measurements by INTEGRAL/ISGRI and Fermi/GBM confirm the overall spin down of the neutron star in GX 1+4 observed over the last decades. This spin down is stronger than ever observed before with an increasement of the spin period of $\sim50\%$ during the last $\sim30$ years reaching the current value of $\sim 160$~s (\cite{gonzalez2010}). The analysis of Fermi/GBM data reveals a negative correlation between spin frequency derivative and the X-ray flux simultaneously measured by Swift/BAT monitor (\cite{gonzalez2010}), which confirms the correlation of instantaneous spin-down torque with X-ray flux discovered by CGRO/BATSE (\cite{chakrabarty97}) discarding clearly the standard accretion disk model for this system. We have shown that in GX 1+4 not only the retrograde disk accretion is a possible explanation but the quasi-spherical accretion onto the neutron star from the stellar wind of the M giant companion is likely to take place (\cite{gonzalez2010}).

\end{document}